\documentstyle[aps]{revtex}
\begin{document}

\vskip 1.truein
\bigskip
\bigskip
\centerline{\Large\bf Theory for polymer coils with necklaces
of micelles}
\vspace{0.5in}

\centerline{\large\bf Richard P. Sear}

\vspace{0.5in}
\centerline{\large FOM Institute for Atomic and Molecular Physics,}
\centerline{\large Kruislaan 407, 1098 SJ Amsterdam, The Netherlands}
\vspace{0.5in}
\centerline{email: sear@amolf.nl}
\vspace{0.5in}

\centerline{\today}
\vspace{1.0in}

\begin{abstract}

If many micelles adsorb onto the same polymer molecule then they
are said to form a necklace. A minimal model of such a necklace
is proposed and shown to be almost equivalent to a 1-dimensional
fluid with nearest-neighbour interactions. The thermodynamic
functions of this fluid are obtained and then used to predict the change
in the critical micellar concentration of the surfactant in
the presence of the polymer. If the amount of polymer is not too large
there are two critical micellar concentrations, one for micelles
in necklaces and one for free micelles.

\end{abstract}

\newpage

\section{Introduction}

Experiment has shown that some micelles adsorb strongly onto polymer chains.
If the polymer molecule
is large enough many micelles can adsorb onto one molecule, forming
a necklace of micelles along the polymer chain.
The most widely studied example
is the adsorption of micelles
of sodium dodecyl sulphate (SDS) onto
poly(ethylene oxide) (PEO), see Refs.
\cite{cabane85,brown92} and references
therein. See Ref. \cite{feitosa96} for work on other systems.
The interaction energy of a micelle with a polymer chain can be strong,
many times $kT$ the thermal energy, and has a large effect on both
the polymer and the micelle. The favourable interaction of a micelle
with the polymer reduces the surfactant density at which the micelles form;
the critical micellar concentration (cmc) is reduced. Whilst if many
micelles adsorb onto a single polymer the repulsive micelle--micelle
interactions cause the polymer chain to swell; the radius of gyration
may increase by a factor of two.
Below, we propose a simple minimal model for necklaces of micelles
adsorbed on a polymer chain. We then go on to calculate the free energy
of such a necklace and use this to determine when they form and to
look at the competition between free micelles and
micelles bound to polymer.

A polymer chain is a linear object,
its topological dimension is 1 \cite{mandelbrot82}. So,
two micelles adsorbed onto a chain
cannot pass each other without one of them detaching from
the polymer chain. Therefore the necklace of micelles along a polymer chain
behaves as a 1-dimensional system.
In the limit that the polymer
molecule is very long the necklace is a 1-d bulk fluid.
We assume that the micelle--micelle interactions
are repulsive and so no two micelles can approach each other closely.
This restriction on the relative positions of the micelles of the necklace
restricts the configurations the polymer can adopt and so reduces its
entropy.
At high density of micelles the micelles are closely spaced
along the polymer, that is the pieces of polymer between the micelles are
quite short, and so these short pieces of
polymer have to stretch in order to
bridge the gaps between the micelles.
This loss of entropy eventually limits the number of micelles which
adsorb on a polymer and when the polymer chains are saturated with
micelles, free micelles are formed.

In the next section we derive from first principles a theory
of objects with excluded volume interactions, which are adsorbed onto
a linear polymer. This is the our main result, our results
are not qualitatively new. They are however, semiquantitative, unlike
previous work which was more qualitative.\cite{nikas94}
First, the polymer-mediated interaction between two micelles
is derived, yielding a potential of mean force for the
intermicellar interaction. Then this is combined with a theory
of 1-dimensional fluids to provide the equation of state of
particles adsorbed onto a polymer. 
This is then used in section III to calculate the equation of state of
a necklace.
Combining the equation of state with a standard theory
of micelle formation \cite{israelachvili76,israelachvili92} then
allows us to calculate how many micelles will form, both absorbed
on polymer and free, as a function of the surfactant density.
Finally, we discuss possible extensions of the theory and model.

\section{Theory for necklaces of micelles}

The system is a mixture of surfactant and polymer in a common solvent;
the mixture is dilute, i.e., the fraction of the volume taken
up by the surfactant and polymer is small.
The polymer is composed of coils of size $L$, where $L$ is proportional
to the number of monomers of a polymer molecule \cite{doi86,eisenriegler93}.
The coils are ideal,
Gaussian, coils and if unperturbed the mean square distance between the
two ends of a coil equals $L$. Thus $L$ has dimensions of length
squared, and $L^{1/2}$ is the single relevant length scale for the
polymer \cite{doi86,eisenriegler93}.
We are assuming that the polymer's Kuhn length is much
smaller than any length we consider explicitly. Our theory is
entirely mesoscopic: the micelle is treated simply as a
sphere and the polymer--micelle interaction is not considered
explicitly but is treated phenomenologically via an association
constant. This constant $\Delta\mu$ is the difference in
excess chemical potential between a micelle bound
to a piece of a polymer chain and a free micelle.
Here, excess means the chemical
potential minus its ideal gas part: $\ln\rho$, where $\rho$ is the
(1- or 3-d) density. If $\Delta\mu$ is negative the micelles will tend to
adsorb on the chain \cite{alexander77}.
This adsorption is due to part of the polymer
chain lying near the surface of the micelle and interacting
via an attractive interaction with it \cite{brown92}. However, for
simplicity we treat the polymer--micelle interaction as that between
the micelle and a point on the polymer; we ignore the amount of polymer
taken up in adsorbing to the micelle. With this simplification a
micelle adsorbed onto a polymer coil is free to move along the
entire length of the polymer $L$. As the polymer is a 1-dimensional (1-d)
chain of monomers the micelle acts as a particle on a wire. Even
though the polymer is tracing out a an extremely
convoluted path through 3-d space the configurational
space available to a micelle adsorbed onto the polymer is only
1-d.

If there is more than one micelle adsorbed on the same polymer coil then
the micelles will interact with each other.
We model the micelle--micelle
interaction via a simple hard-sphere interaction: no two micelles'
centres may be within a distance $D$ of each other.
The size of the micelles, $D$, is much less than that of a coil, $L^{1/2}$,
allowing many micelles to adsorb onto one coil.
Above, we assumed that the polymer--micelle adsorption was pointlike, i.e.,
a negligible amount of the polymer is adsorbs on a micelle and we
have just assumed that two micelles interact with each other out to
a distance $D$. For these two assumptions to be consistent the
physical size of the micelles must
be much smaller than the range at which two micelles interact.
Essentially, we must have a micelle of diameter $\ll D$ which is
quite highly charged so that no other micelle can approach within
$D$ \cite{israelachvili92,cabane87}. As micelles are typically
3--4 nm in diameter, $D\gg 4$nm.

The limit of few micelles adsorbed onto a polymer chain, when
micelle--micelle interactions are unimportant,
is trivial (at our phenomenological level of description). It is
just the free energy of a polymer coil plus the excess free energy
of a free micelle plus
$\Delta\mu$ plus the log of the 1-d density of micelles on the chain.
The first of these is a reference free energy,
which we need not specify, and the free energy of a free micelle and
$\Delta\mu$ are parameters of our model.
However, at higher densities micelles on the same coil interact.
They do so when the micelles are close together, i.e., when the
pieces of polymer which connect adjacent micelles are short.
The partition function of a piece of polymer chain with a micelle
at each end is different from the partition function of the same length
of chain without the micelles.
In evaluating the
partition function when the micelles are present,
we must exclude those configurations in which the
two micelles' hard spheres overlap. This is done in the
following section and in section IIA we use that result to
calculate the thermodynamic functions for a necklace of micelles
in the limit that the coil is infinitely long.
In IIB this result is used to determine free energy of a necklace
at any density.

\subsection{The interaction between two micelles adsorbed onto
the same polymer coil}

We consider two micelles adsorbed onto a coil and separated by a length
of polymer $L_{12}$.
The probability of this separation of the micelles is proportional to
the partition function of the chain at this separation.
The two micelles have split the coil into 3 pieces, the one between them
of length $L_{12}$ and the two end pieces. The three parts are
independent as the coil is ideal. Only the partition function of the
centre piece is affected by the micelles; the end pieces
are free to adopt any configuration but the the ends of the centre piece
are each attached to a micelle and so, because of the micelle--micelle
interaction, they must be at least $D$ apart. The partition function
of a chain of length $L_{12}$ whose ends are constrained to be
at least $D$ apart is easy to calculate.
The partition function for a chain
with ends separated by ${\bf r}$ divided by its partition function
when free, is just the probability distribution
function $\phi$ for the ends of an unconstrained chain, times $d{\bf r}$.
Now, $\phi$ is
\cite{doi86}
\begin{equation}
\phi(L_{12},{\bf r})=\left(\frac{3}{2\pi L_{12}}\right)^{3/2}
\exp\left(\frac{-3r^2}{2L_{12}}\right).
\end{equation}
Thus, the partition function with the ends at least $D$ apart,
divided by the partition function for an unconstrained chain is simply
\begin{equation}
4\pi\int_D^{\infty}\phi(L_{12},{\bf r})r^2dr
=1-\frac{4}{\pi^{1/2}}\int_0
^{\left(\frac{3}{2}\right)^{1/2}D/L_{12}^{1/2}}
x^2\exp(-x^2)dx=\exp[-w(L_{12}/D^2)].
\label{wdef}
\end{equation}
The seond equality defines $w$ the effective interaction potential
between two neighbouring micelles.
The probability of the micelles being $L_{12}$ apart, $P(L_{12}/D^2)$,
is directly related to the partition function of the chain when the
micelles are $L_{12}$ apart,
\begin{equation}
P(L_{12}/D^2)=
\frac{\exp[-w(L_{12}/D^2)]}{\int\exp[-w(L_{12}/D^2)]dL_{12}},
\end{equation}
for an infinitely long or ring polymer.
We see that $\exp[-w(L_{12}/D^2)]$ is acting as
a Boltzmann weight and so $w$ is an effective interaction;
it is plotted in Fig. (\ref{fig2}).
The interaction $w$ is an athermal but soft repulsion.
When two micelles are bound to the polymer
less then $L_{12}\sim D$ apart along a polymer
chain then the chain has to stretch to cross the distance between
the micelles, which must be $\ge D$.
Conventionally, the Boltzmann weight is the exponential of
minus an energy of interaction but here it is the exponential of
minus the free energy of the piece of polymer coil. However, despite
its unusual origin $w$ behaves just as an interaction energy.
The pair of micelles
behave as a pair of particles on a wire which interact via a
potential $w$. In the following section we use $w$ to obtain the
thermodynamic functions of a fluid of many micelles on one polymer coil.

\subsection{A 1-dimensional fluid of micelles on a polymer coil}

When many micelles are adsorbed onto the same coil they behave as
a 1-d fluid. Formally we will take the $L/D^2\rightarrow\infty$
limit, producing an infinite fluid. We do not consider effects
due to the finite size of the polymer coils.
Each micelle's movement is restricted by
the micelle in front and the micelle behind. As the system is
effectively 1-d the micelles form a fluid at all densities;
they never solidify or undergo a liquid--vapour transition
\cite{landau,lieb66}. Due to the convoluted path taken by the chain
non-adjacent micelles can interact, indeed this interaction will
cause the coil to swell and change from being Gaussian to being swollen
\cite{doi86,degennes}. However, these interactions are weak in
comparison to the interactions between nearest neighbour micelles and
so will be neglected for the purpose of determining the free energy
of the necklace of micelles. Each micelle always interacts with
the micelles in front and behind it but collisions between different
parts of the chain are relatively infrequent \cite{degennes}.

Therefore we have a simple 1-d fluid. As the interaction between
micelles is mediated via the polymer chain the interaction is
strictly nearest neighbour. The free energy of the coil is simply
a sum of the free energies of the pieces between the micelles and so
is just a function of all the $L_{12}$'s between neighbouring micelles.
The statistical mechanics of 1-d fluids with nearest neighbour
interactions is both well-studied and simple \cite{lieb66}.
We start from the canonical
partition function $Z_N(L)$ of $N$ micelles on a coil of length $L$.
As usual for 1-d systems, molecules fixed at each end of the system
define its limits. So, with a micelle fixed at coordinate
$x_0$ and one at $x_N$, $Z_N$ is 
\begin{equation}
Z_N(L)=\int_{x_i<x_{i+1}}\prod_{i=1}^Ndx_i
\exp\left[-\sum_{i=1}^{N+1}w(x_i-x_{i-1})\right],
\end{equation}
where the factor of $1/N!$ is absent because we have put the
micelles into a sequence: the $i$'th micelle is always the
left-hand neighbour of the $(i+1)$'th micelle \cite{lieb66}.
The units are defined so that the thermal energy $kT$ is unity.
Progress can be made if we move to the isothermal-isobaric
ensemble \cite{lieb66,hansen86}; introducing the pressure $p$ which
is the conjugate variable to the length $L$.
The isothermal-isobaric partition function $Z_N(p)$ is
\begin{equation}
Z_N(p)=\int Z_N(L)\exp(-pL)dL.
\end{equation}
Following the method of Takahashi \cite{lieb66} we note that as
the $(N+1)$'th particle defines the limit of our system $x_{N+1}=L$.
Then if we define the new coordinates $r_i=x_i-x_{i-1}$,
\begin{eqnarray}
Z_N(p)&=&\int_{r_i>0}\prod_{i=1}^{N+1}dr_i
\exp\left[-\sum_{i=1}^{N+1}\left\{w(r_i)+pr_i\right\}\right]\nonumber\\
&=&\left(\int_0^{\infty}dr\exp[-w(r)-pr]\right)^{N+1},
\label{zp}
\end{eqnarray}
the partition function has reduced to a simple 1-d integration.
The chemical potential $\mu_{1d}$ is, substituting $N$ for $N+1$
in Eq. (\ref{zp}),
\begin{equation}
\mu_{1d}=
-\frac{1}{N}\ln Z_N(p)=-\ln \int_0^{\infty}dr\exp[-w(r)-pr],
\label{mu1d}
\end{equation}
which gives $\mu_{1d}$ as a function of the pressure $p$.
We would like it as a function of the reduced 1-d density
$\rho_{1d}=ND^2/L$.
The density $\rho_{1d}$ is obtained from
\begin{equation}
\rho_{1d}=\left(\frac{\partial\mu_{1d}}{\partial pD^2}\right)^{-1}.
\label{rho1d}
\end{equation}
Using Eq. (\ref{wdef}) in Eq. (\ref{mu1d}) and then Eq. (\ref{rho1d})
yields $\mu_{1d}$ as a function of density $\rho_{1d}$; it is
plotted in Fig. (\ref{fig3}). Notice that the chemical potential
is almost a linear function of density at high density.
For future use we define an excess chemical potential
$\mu_{ex}=\mu_{1d}-\ln\rho_{1d}$.
We now possess the chemical potential of a necklace of micelles as
a function of the density of micelles along the polymer chain. In
order to determine the density of micelles along a polymer
as a function of surfactant density
we require a model for the formation of micelles.

\section{Micelle formation}

Our model for micelle formation is quite standard
\cite{israelachvili76,israelachvili92}. We start by assuming that
the surfactant monomers, the free micelles, and the micelles
bound to polymer coils form an ideal ternary mixture.
The only interactions considered are those
between micelles bound
to the same polymer; interactions involving free micelles or monomers
are neglected as the surfactant + polymer solution is dilute.
The ternary mixture's density and composition are specified by
three densities: the density of surfactant monomers
$\rho_0=N_0D^3/V$, of free micelles $\rho_{fm}=N_{fm}D^3/V$, and of
bound micelles $\rho_{bm}=N_{bm}D^3/V$.
$N_0$, $N_{fm}$ and $N_{bm}$ are the numbers of surfactant
monomers, free micelles and micelles bound to polymer, respectively, and
$V$ is the volume.
The 1-d density of micelles along the polymer chains $\rho_{1d}$ is related
to the density of bound micelles in the mixture $\rho_{bm}$ by
\begin{equation}
\rho_{bm}=\rho_p\rho_{1d},
\label{1d3d}
\end{equation}
where
$\rho_p=N_pLD/V$ is a `segment' density of the polymer.
It is the number density of segments of polymer of size $D^2$, times $D^3$.
$N_p$ is the number of
polymer coils.
Below, we treat the monomers and free micelles, and the micelles bound
to polymer as constituting two different bulk phases.
The polymer density $\rho_p$
just defines the relative volumes of these two parts of our system.

At equilibrium the chemical potential of
a surfactant is the same if it is a monomer or is part of either
type of micelle.
Then the chemical potential of a micelle composed of $m$ surfactant
molecules is $m$ times that of a monomer. For
the chemical potential of a monomer to be equal to that in a free
micelle \cite{israelachvili76,israelachvili92}
\begin{equation}
m(\ln\rho_0+\mu_0)=\ln\rho_{fm}+\mu_{mic},
\label{mueq1}
\end{equation}
and for it to equal that in a bound micelle
\begin{equation}
m(\ln\rho_0+\mu_0)=
\mu_{1d}(\rho_{1d})+\mu_{mic}+\Delta\mu=
\mu_{1d}(\rho_{bm}/\rho_p)+\mu_{mic}+\Delta\mu,
\label{mueq2}
\end{equation}
where we used Eq. (\ref{1d3d}) to go from the middle to the right
hand side expression.
We have assumed that the
number of surfactant molecules in a micelle is $m$ for both free
and bound micelles.
Micelles are of course not monodisperse and so this is an approximation,
although the numbers of micelles
much larger or smaller than the most probable size is small
\cite{israelachvili76,israelachvili92}.
The constants $\mu_0$ and $\mu_{mic}$ are the contributions
to the chemical potentials of the monomers and micelles, respectively,
of the interaction with the solvent. They are the change in free energy
in excess of the $\ln\rho$ term,
of the system when a monomer or a micelle is inserted
\cite{israelachvili92}.
We have neglected the contribution from the
momentum degrees of freedom.
Thus, $\mu_{mic}-m\mu_0$ is the difference
between the chemical potential (minus the $\ln\rho$ terms) of
$m$ surfactant molecules in a micelle and $m$ surfactant monomers.

The first equation for equilibrium between monomers and micelles,
Eq. (\ref{mueq1}), is
easily rewritten in the familiar form
\begin{equation}
\rho_{fm}=\rho_0^m\exp(m\mu_0-\mu_{mic}).
\label{rhofm}
\end{equation}
The most convenient thermodynamic variable to work in is $\rho_0$ which
corresponds to controlling the chemical potential of the
surfactant: $\ln\rho_0+\mu_0$.
The second equation for equilibrium, Eq. (\ref{mueq2}),
cannot be rewritten in the form
$\rho_{bm}$ equals a function of $\rho_0$ because the
the right hand side includes $\mu_{1d}$ which is not a simple
logarithmic or linear function of $\rho_{bm}$.
If $\rho_p$, $\mu_{mic}$ and
$\Delta\mu$ are specified, then Eq. (\ref{mueq2}) can be solved as a
non-linear equation for $\rho_{bm}$, at any value of $\rho_0$.

The total number density of surfactant is proportional to
$\rho=\rho_0/m+\rho_{fm}+\rho_{bm}$. As all three of the densities
$\rho_0$, $\rho_{fm}$ and $\rho_{bm})$ are made dimensionless by
multiplying by $D^3$, the volume fraction of the surfactant is
approximately equal to $\rho$ times the physical volume of a micelle
divided by $D^3$. The physical volume of a micelle is the volume actually
occupied by the surfactant molecules. As the physical volume is much less
than $D^3$ the surfactant's volume fraction is much less than $\rho$. 
In Figs. (\ref{fig4}) and (\ref{fig5}) we plot the
densities of monomers and micelles as a function of this total
surfactant density. In Fig. (\ref{fig4})
we have no polymer and there are only free micelles.
The critical micellar concentration (cmc) \cite{israelachvili92}
is clearly around $10^{-2}$, although it is not that well defined
as our aggregation number $m$ is not that large, $m=20$.
In Fig. (\ref{fig5}), the value of $\mu_{mic}$ is unchanged
from Fig. (\ref{fig4}) but there is polymer
present. As $\Delta\mu$ is negative the polymer stabilises the micelles.
This has two effects: micelles form at lower densities,
the polymer encourages micelle formation, and micelles
prefer to be adsorbed onto a polymer than to be free. So, at low
densities where there are few micelles they are nearly all adsorbed onto
polymer chains; the density of free micelles is very low below
$\rho\simeq3\times10^{-2}$. As the density of micelles increases the
density of micelles on the polymer chains becomes quite high and
the interaction free energy of these micelles, $\mu_{ex}$, is large.
The polymer chains are stretching to accommodate the micelles and this
stretching is reducing the configurational entropy of the polymer.
This reduces the free energy reduction on adsorption and so we see
that free micelles start to form.
From Eqs. (\ref{mueq1}) and (\ref{mueq2}),
it is easy to see that when $\mu_{ex}=\ln\rho_p-\Delta\mu$ the numbers
of free and bound micelles are the same.

Notice that the density of surfactant
at which free micelles appear is much higher in Fig. (\ref{fig5})
than in Fig (\ref{fig4}) because the monomer density is depleted by
the formation of bound micelles. Free micelles appear at a
$\rho_0$ determined solely by $\mu_{mic}$, see Eq. (\ref{rhofm}),
but this corresponds to
a higher $\rho$ if there is competition for monomers.
The $\rho_0$ curve of Fig. (\ref{fig5}) looks as if there are two cmc's;
it has two steps while in Fig. (\ref{fig4}) it has only one.
The first is where bound micelles form and the rate
of increase of $\rho_0$ decreases sharply. Then as the density
of bound micelles increases the polymer chains become crowded,
micelle formation becomes less favourable, and the rate of increase
of $\rho_0$ increases. At yet higher density, free micelles form and
the rate of increase of $\rho_0$ decreases again.
Of course, if the polymer density $\rho_p$ increases then the polymer
chains saturate at higher surfactant densities and the second `cmc'
is pushed to higher values of $\rho$.

The formation of a necklace of micelles causes the ideal polymer coil
to expand for two reasons: it stretches the pieces of the polymer chain
between the micelles and the micelle--micelle interaction results
in the chain being swollen, like a polymer coil in a good
solvent \cite{doi86,degennes}. Without micelles the radius of
gyration of the polymer $R_G=L^{1/2}/\sqrt6$. At low densities
of micelles, $\rho_{1d}\lesssim1$ the radius of gyration is
difficult to estimate as the micelles cover only a fraction of the
chain. Therefore some parts, with micelles, repel each other but
other, bare, parts do not. However, for $\rho_{1d}>1$ almost
all the chain is covered with micelles. In order to minimise the
stretching of the polymer the spacing between the micelles will be
quite regular and near $D$. So, we can consider the chain
as being composed of $\rho_{1d}L/D^2$ segments of length $\sim D$.
Then as the chain is swollen $R_G\sim (\rho_{1d}L/D^2)^{3/5}D$.
As $L\gg D^2$ this is much larger than $L^{1/2}$.

\section{Conclusion}

Perhaps the simplest possible model of polymer coils with necklaces
of micelles has been proposed and studied. The simplicity of the model
has allowed a detailed examination of its behaviour with only very
minor approximations. This differs from previous theoretical
work \cite{nikas94} which involved a highly approximate treatment
of a much more detailed, and so complex, model.
We have shown that the 1-d topology
\cite{mandelbrot82} of a polymer
chain implies that the necklace behaves as a 1-d fluid. This allowed us
to determine the behaviour of the necklace using the highly
developed theory for 1-d systems \cite{lieb66}.
In combination with a standard theory for micelle formation it was then
straightforward to calculate the densities of micelles formed,
see Fig. (\ref{fig5}).

The model was kept simple for the sake of simplicity and clarity.
For the purposes of comparison with experiment a more complex and
accurate model is required. The theory presented here can be extended
to deal with the most unrealistic features of the model.
For example, the theory neglects the finite size of the polymer
coils. Micelles are quite large, $\sim 4$nm across, and so if the polymer
coil is small there may be room for only a few micelles to adsorb
without stretching the polymer so much that it is not favourable to adsorb
another micelle. Then the fluid of micelles is not a bulk fluid.
The finite
size of the coil corresponds to a 1-d fluid between two walls, $L$ apart.
Confined 1-d fluids have been considered extensively, see Ref.
\cite{percus82}.
The coils studied
in experiment \cite{brown92}
are in a good solvent and so are swollen.
Therefore it would be worthwhile to extend our theory for ideal coils to
self-avoiding coils.
Although self-avoiding coils are more complex and
difficult to deal with than ideal coils \cite{doi86,degennes}, enough
is known about their free energy when they are stretched \cite{degennes}
to perform at least a scaling theory of a polymer with a micellar necklace.
Experimental
results \cite{cabane87} on a semidilute solution of polymer
clearly show the competition between polymer
entropy and the free energy of adsorption which we have observed here.
We hope that the theory
developed here can be applied even to this system.

It is a pleasure to acknowledge
a careful reading of the manuscript by J. Doye    .
I would like to thank The Royal Society for the award of a fellowship
and the FOM institute AMOLF for its hospitality.
The work of the FOM Institute is part of the research program of FOM
and is made possible by financial support from the
Netherlands Organisation for Scientific Research (NWO).

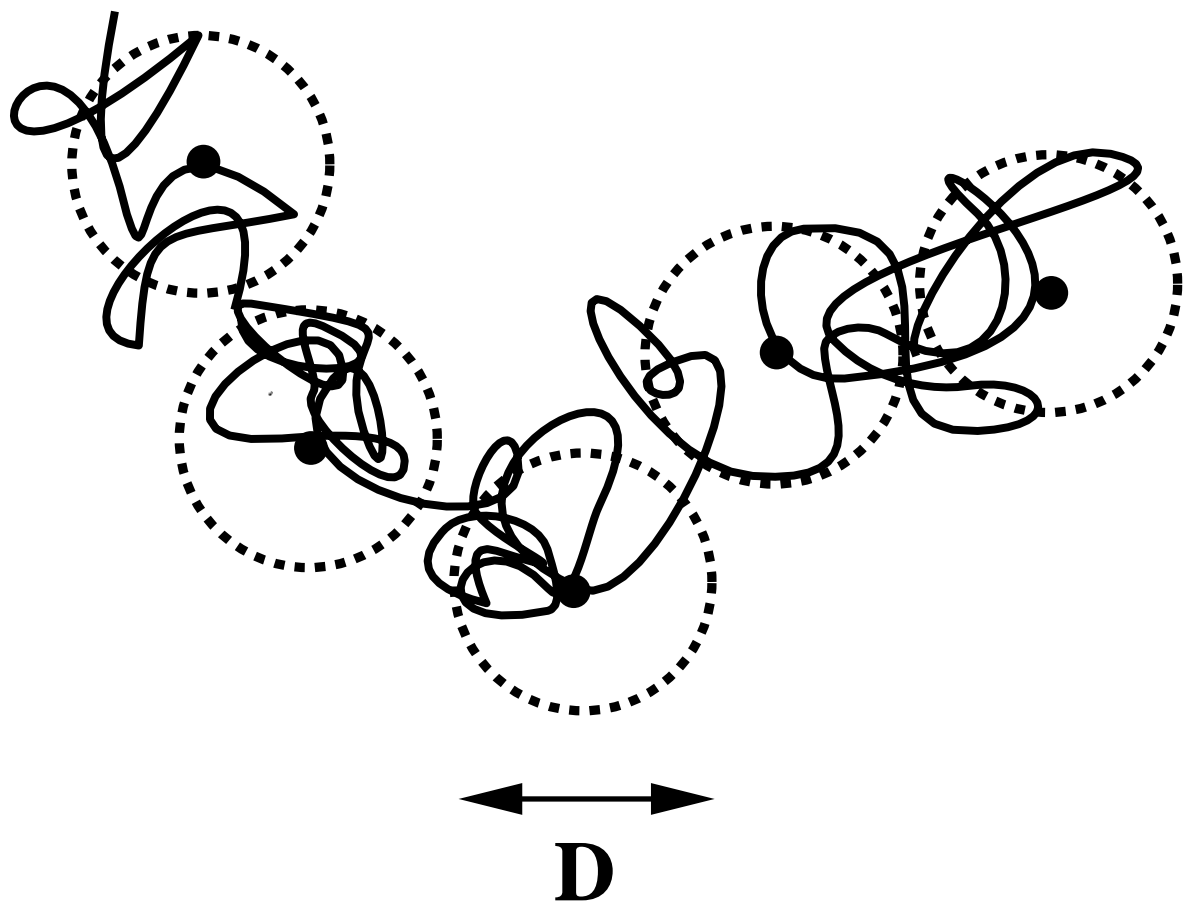
\begin{figure}
\caption{
A schematic picture of a polymer coil with a necklace of micelles.
The polymer chain is the black curve and the micelles adsorbed onto the
chain are represented by dotted circles of diameter $D$.
}
\label{fig1}
\end{figure}

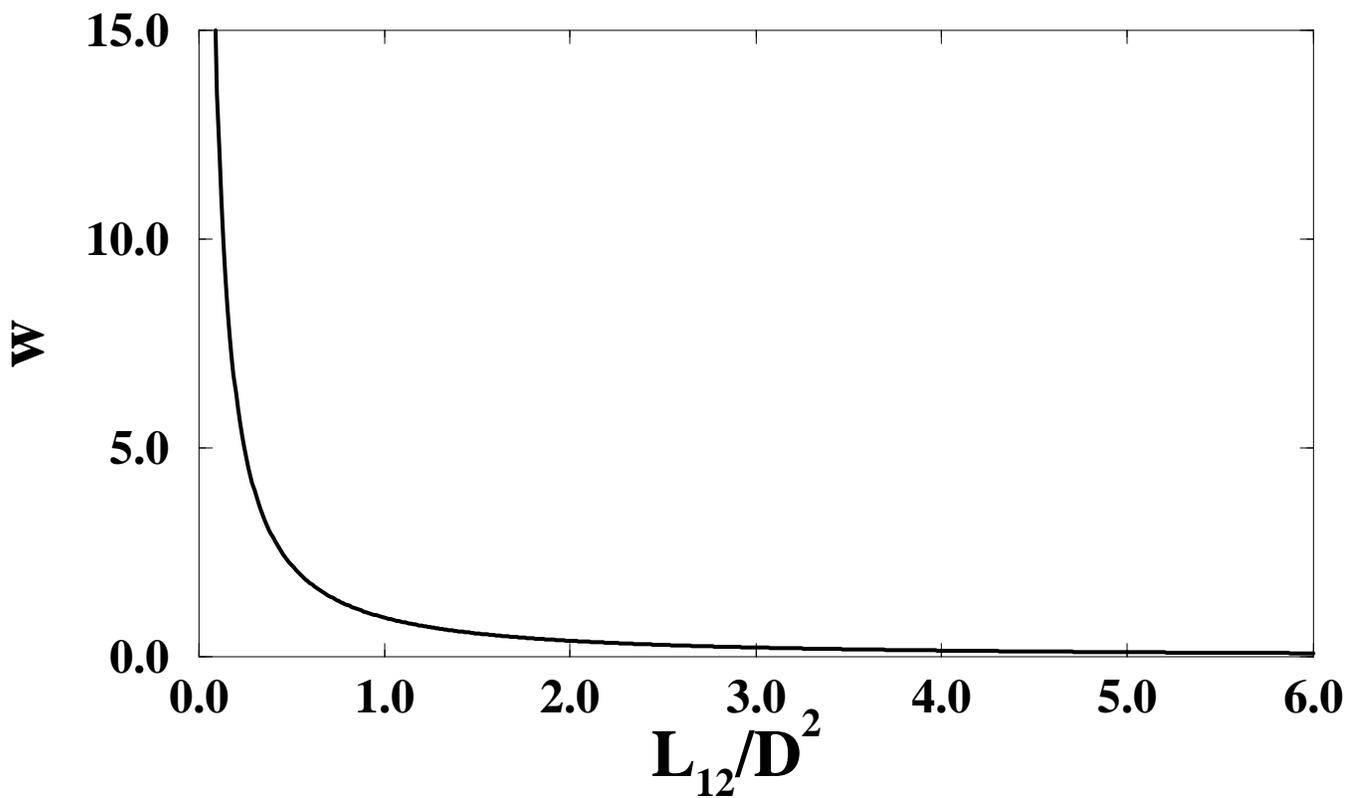
\begin{figure}
\caption{
The polymer mediated, effective interaction $w$ between two micelles
adsorbed onto a polymer coil, as a function of their separation $L_{12}$.
}
\label{fig2}
\end{figure}

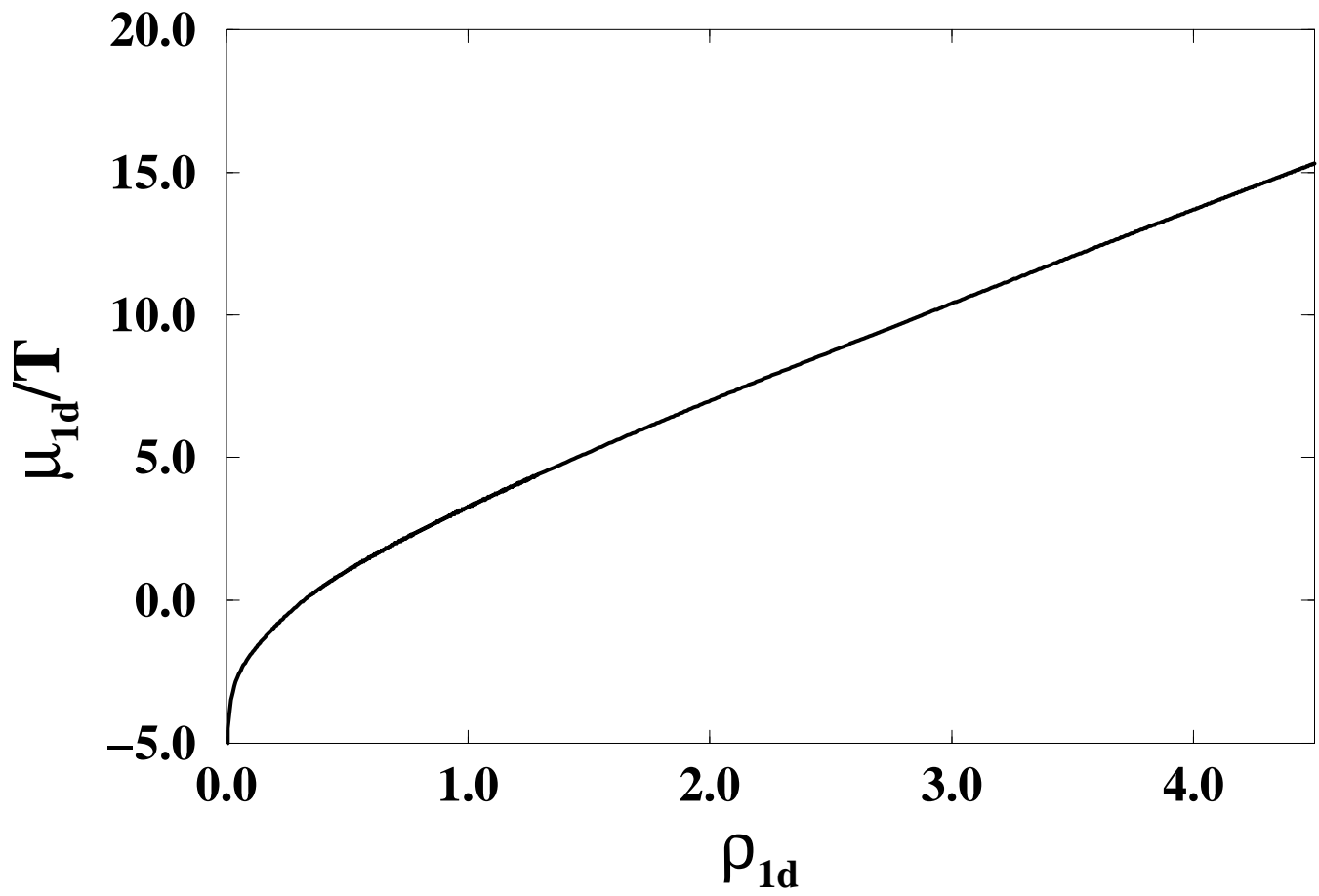
\begin{figure}
\caption{
The chemical potential of a 1-d fluid of micelles on a polymer
coil, interacting via the potential $w$.
}
\label{fig3}
\end{figure}

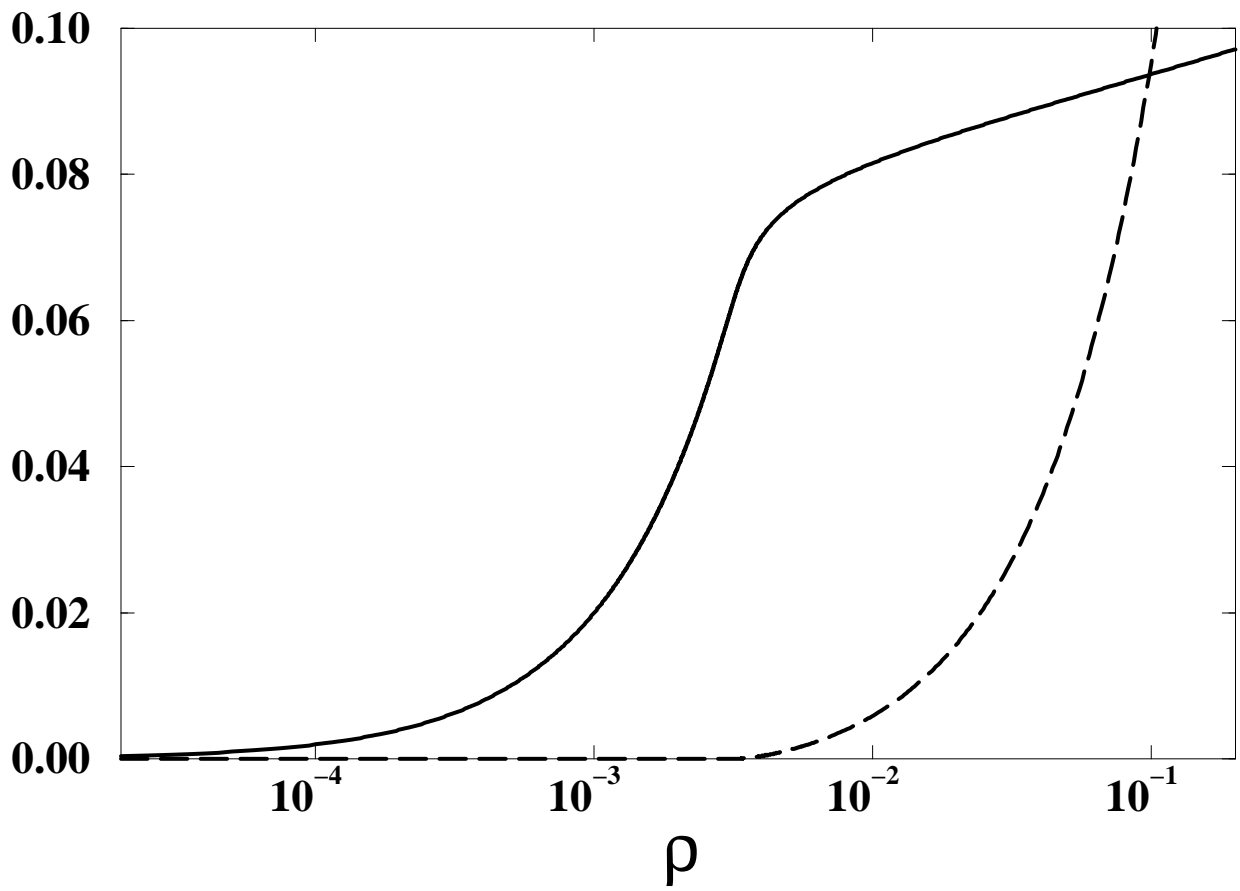
\begin{figure}
\caption{
The density of surfactant monomers $\rho_0$ (solid curve), and
of free micelles $\rho_{fm}$ (dashed curve) as a function of total
surfactant density $\rho$. $m\mu_0-\mu_{mic}=45$ and $m=20$.
There is no polymer present.
}
\label{fig4}
\end{figure}

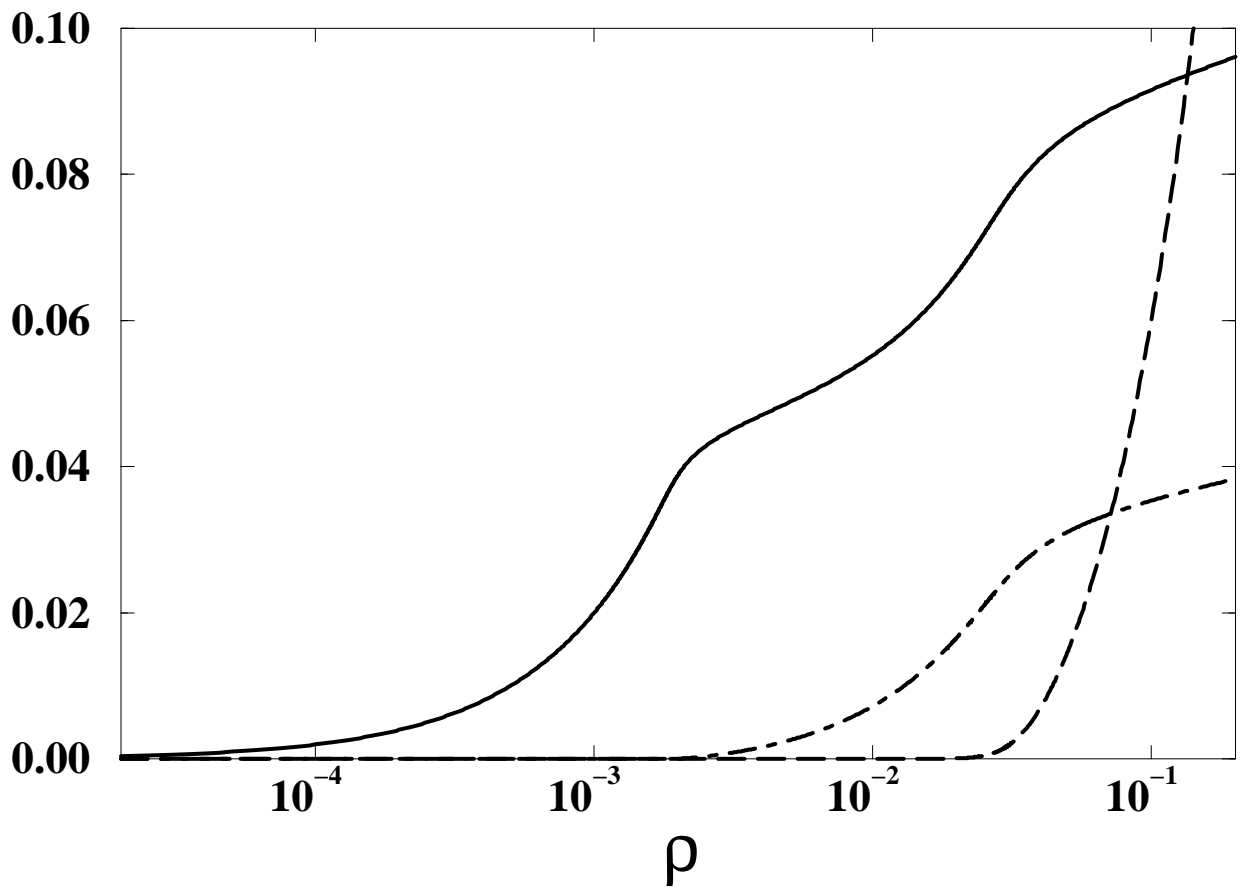
\begin{figure}
\caption{
The density of surfactant monomers $\rho_0$ (solid curve),
of free micelles $\rho_{fm}$ (dashed curve), and
of bound micelles $\rho_{fm}$ (dot-dashed curve) as a function of total
surfactant density $\rho$. $m\mu_0-\mu_{mic}=45$, $m=20$,
$\Delta\mu=-15$ and $\rho_p=0.01$.
}
\label{fig5}
\end{figure}
\end{document}